\documentclass[runningheads]{llncs}
\usepackage{graphicx}
\usepackage{times}
\usepackage{epsfig}
\usepackage{graphicx}
\usepackage{amssymb}
\usepackage{pifont}
\usepackage{amsmath}
\usepackage{amssymb}
\usepackage{dsfont}
\newcommand{\app}{\raise.17ex\hbox{$\scriptstyle\sim$}}

\begin{document}
\title{Hyper-Pairing Network for Multi-Phase Pancreatic Ductal Adenocarcinoma Segmentation}
\titlerunning{HPN for Multi-Phase Pancreatic Ductal Adenocarcinoma Segmentation}
%
\author{Yuyin Zhou\inst{1} \and
Yingwei Li\inst{1} \and
Zhishuai Zhang\inst{1} \and
Yan Wang\inst{1} \and
Angtian Wang\inst{2} \and
Elliot K. Fishman\inst{3} \and
Alan L. Yuille\inst{1} \and
Seyoun Park\inst{3}}
\authorrunning{Y. Zhou et al.}

\institute{The Johns Hopkins University, Baltimore, MD 21218, USA \and
Huazhong University of Science and Technology, Wuhan 430074, China \and
The Johns Hopkins University School of Medicine, Baltimore, MD 21287, USA
}

\maketitle              
\begin{abstract}
Pancreatic ductal adenocarcinoma (PDAC) is one of the most lethal cancers with an overall five-year survival rate of $8\%$. Due to subtle texture changes of PDAC, pancreatic dual-phase imaging is recommended for better diagnosis of pancreatic disease.
In this study, we aim at enhancing PDAC automatic segmentation by integrating multi-phase information (\emph{i.e.}, arterial phase and venous phase). 
To this end, we present Hyper-Pairing Network (HPN), a 3D fully convolution neural network which effectively integrates information from different phases. The proposed approach consists of a dual path network where the two parallel streams are interconnected with hyper-connections for intensive information exchange.
Additionally, a pairing loss is added to encourage the commonality between high-level feature representations of different phases. Compared to prior arts which use single phase data, HPN reports a significant improvement up to $7.73\%$ (from $56.21\%$ to $63.94\%$) in terms of DSC.

\end{abstract}
\begin{figure}[t]
\centering
\includegraphics[width=0.89\linewidth]{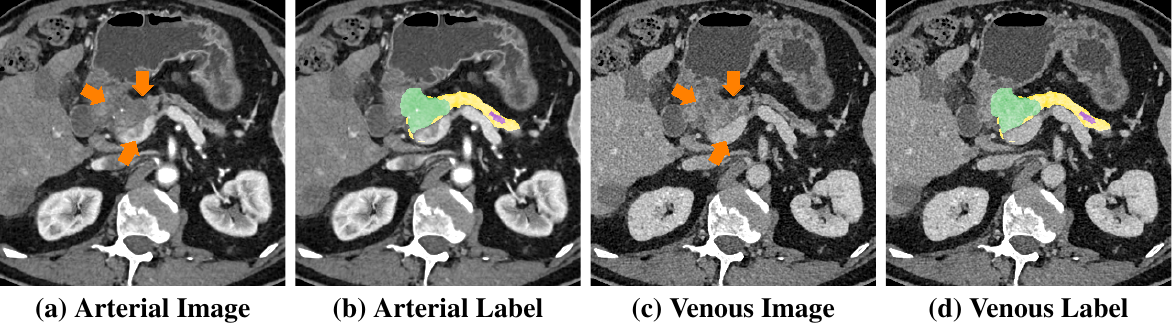}
\caption{Visual comparison of arterial and venous images (after alignment) as well as the manual segmentation of normal pancreas tissues (yellow), pancreatic duct (purple) and PDAC mass (green). Orange arrows indicate the ambiguous boundaries and differences of the abnormal appearances between the two phases. Best viewed in color.}

\label{fig:main_fig}
\end{figure}

\section{Introduction}
\label{Introduction}

Pancreatic ductal adenocarcinoma (PDAC) is the 4th most common cancer of death with an overall five-year survival rate of 8\%.
Currently, detection or segmentation at localized disease stage followed by complete resection can offer the best chance of survival, \emph{i.e.}, with a 5-year survival rate of 32\%. The accurate segmentation of PDAC mass is also important for further quantitative analysis, \emph{e.g.}, survival prediction~\cite{attiyeh2018survival}.
Computed tomography (CT) is the most commonly used imaging modality for the initial evaluation of PDAC. However, textures of PDAC on CT are very subtle (Fig.~\ref{fig:main_fig}) and therefore can be easily neglected by even experienced radiologists. 
To our best knowledge, the state-of-the-art on this matter is \cite{zhu20173d}, which reports an average Dice of $56.46\%$.
For better detection of PDAC mass, dual-phase pancreas protocol using contrast-enhanced CT imaging, which
is comprised of arterial and venous phases with intravenous contrast delay, are recommended. 

In recent years, deep learning has largely advanced the field of computer-aided diagnosis (CAD), especially in the field of biomedical image segmentation~\cite{dou2016automatic,Ronneberger_2015_UNet,Roth_2016_Spatial,zhou2017fixed,zhu2019anatomynet,wang2019abdominal}.
However, there are several challenges for applying existing segmentation algorithms to dual-phase images. First, segmentation of pancreatic lesion, \emph{e.g.}, cysts~\cite{zhou2017deep}, is more difficult than organ segmentation due to its smaller sizes, lower contrast and texture similarity,~\emph{etc}.
Secondly, these algorithms are optimized for segmenting only one type of input, and therefore cannot be directly applied to handle multi-phase data. More importantly, how to properly handle the variations between different views requires a smart information exchange strategy between different phases. 
While how to efficiently integrate information from multi-modalities has been widely studied~\cite{dolz2018hyperdense,li2019multimodal,zhang2015deep}, the direction on learning multi-phase information has been rarely explored, especially for tumor detection and segmentation purposes.


To address these challenges, we propose a multi-phase segmentation algorithm, Hyper-Pairing Network (HPN), to enhance the segmentation performance especially for pancreatic abnormality. Following HyperDenseNet~\cite{dolz2018hyperdense} which is effective on multi-modal image segmentation, we construct a dual-path network for handling multi-phase data, where each path is intended for one phase. 
To enable information exchange between different phases, we apply skip connections across different paths of the network~\cite{dolz2018hyperdense}, referred as \emph{hyper-connections}. 
Moreover, noticing that a standard segmentation loss (cross-entropy loss, Dice loss~\cite{Milletari_2016_VNet}) only aims at minimizing the differences between the final prediction and the groundtruth thus cannot well handle the variance between different views, we introduce an additional \emph{pairing loss term} to encourage the commonality between high-level features across both phases for better incorporation of multi-phase information.
We exploit three structures together in HPN including PDAC mass, normal panreatic tissues, and pancreatic duct, which serves as an important clue for localizing PDAC. Extensive experiments demonstrate that the proposed HPN significantly outperforms prior arts by a large margin on all 3 targets.


\section{Methodology}
\label{method}

 We hereby focus on dual-phase inputs while our approach can be generalized to multi-phase scans. With phase A and aligned phase B by the deformable registration, we have the set $\mathcal{S} = \{\left(\mathbf{X}_{i}^\textup{A}, \mathbf{X}_{i}^\textup{B}, \mathbf{Y}_{i}\right)|i=1,...,M\}$, where $\textup{X}^\textup{A}_i\in{\mathbb{R}^{W_i\times H_i\times L_i}}$ is  the $i$-th 3D volumetric CT images of phase A with the dimension $\left(W_i\times H_i\times L_i\right)= \mathcal{D}_{i}$ and $\textup{X}^\textup{B}_i\in{\mathbb{R}}^{\mathcal{D}_{i}}$ is the corresponding aligned volume of phase B. $\mathbf{Y}_i = \{ y_{ij} | j=1,..., \mathcal{D}_{i}\}$ denotes the corresponding voxel-wise label map of the $i$-th volume, where $y_{ij}\in\mathcal{L}$ is the label of the $j$-th voxel in the $i$-th image, and $\mathcal{L}$ denotes the label of the target structures. In this study, $\mathcal{L}$=\{normal panreatic tissues, PDAC mass, pancreatic duct\}. The goal is to learn a model to predict label of each voxel $\hat{\mathbf{Y}}={f(\mathbf{X}^\textup{A}, \mathbf{X}^\textup{B})}$ by utilizing multi-phase information.

\begin{figure}[t]
\centering
\includegraphics[width=0.78\linewidth]{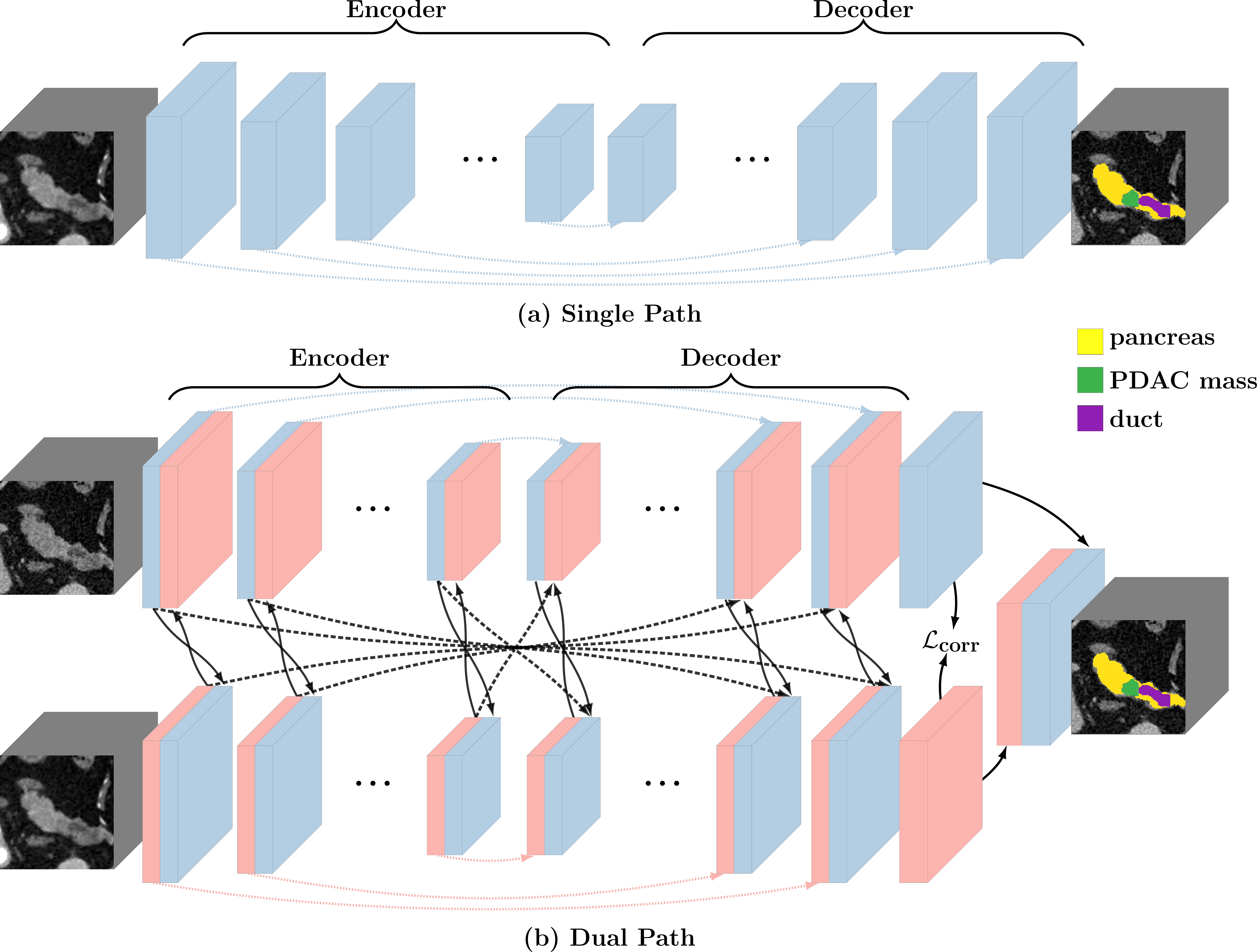}

\caption{(a) The single path network where only one phase is used. The dash arrows denote skip connections between low-level features and high-level features. (b) HPN structure where multiple phases are used. 
The black arrows between the two single path networks indicate hyper-connections between the two streams. An additional pairing loss is employed to regularize view variations, therefore can benefit the integration between different phases. Blue and pink stand for arterial and venous phase, respectively.}

\label{fig:dual_path}
\end{figure}

\subsection{Hyper-connections}
\label{baseline2}
Segmentation networks (\emph{e.g.}, UNet~\cite{Ronneberger_2015_UNet,cciccek20163d}, FCN~\cite{Long_2015_Fully}) usually contain a contracting encoder part and a successive expanding
decoder part to produce a full-resolution segmentation result as illustrated in Fig.~\ref{fig:dual_path}(a). As the layer goes deeper, the output features evolve from low-level detailed representations to high-level abstract semantic representations. The encoder part and the decoder part share an equal number of resolution steps~\cite{Ronneberger_2015_UNet,cciccek20163d}.

However, this type of network can only handle single-phase data.
We construct a dual path network where each phase has a branch with a U-shape encoder-decoder architecture as mentioned above. These two branches are connected via hyper-connections which enrich feature representations by learning more complex combinations between the two phases. Specifically, hyper-connections are applied between layers which output feature maps of the same resolution across different paths as illustrated in Fig.~\ref{fig:dual_path}(b). Let $\textbf{R}_{1}, \textbf{R}_{2},..., \textbf{R}_{\textup{T}}$ denote the intermediate feature maps of a general segmentation network, where $\textbf{R}_{t}$ and $\textbf{R}_{\textup{T} - t}$ share the same resolution ($\textbf{R}_{t}$ is on the encoder path and $\textbf{R}_{\textup{T} - t}$ is on the decoder path).
Hyper-connections are applied as follows: $\textbf{R}^\textup{A}_{t}\longrightarrow \textbf{R}^\textup{B}_{\textup{t}}$, $\textbf{R}^\textup{B}_{t}\longrightarrow \textbf{R}^\textup{A}_{\textup{t}}$, $\textbf{R}^\textup{A}_{t}\longrightarrow \textbf{R}^\textup{B}_{\textup{T} - t}$, $\textbf{R}^\textup{B}_{t}\longrightarrow \textbf{R}^\textup{A}_{\textup{T} - t}$,$\textbf{R}^\textup{A}_{\textup{T} - t}\longrightarrow \textbf{R}^\textup{B}_{\textup{T} - t}$, $\textbf{R}^\textup{B}_{\textup{T} - t}\longrightarrow \textbf{R}^\textup{A}_{\textup{T} - t}$, while maintaining the original skip connections that already occur within the same path, \emph{i.e.}, $\textbf{R}^\textup{A}_{t}\longrightarrow \textbf{R}^\textup{A}_{\textup{T} - t}$, $\textbf{R}^\textup{B}_{t}\longrightarrow \textbf{R}^\textup{B}_{\textup{T} - t}$.





\subsection{Pairing loss}
\label{loss}
The standard loss for segmentation networks only aims at minimizing the difference between the groundtruth and the final estimation, which cannot well handle the variance between different views. Applying this loss alone is inferior in our situation since the training process involves heavy integration of both arterial information and venous information. 
To this end, we propose to apply an additional
pairing loss, which encourages the commonality between the two sets of high-level semantic representations, to reduce view divergence. 

We instantiate this additional objective as a correlation loss~\cite{yao2017deep}. Mathematically, for any pair of aligned images ($\textup{X}^\textup{A}_i$, $\textup{X}^\textup{B}_i$) passing through the corresponding view sub-network, the two sets of high-level semantic representations (feature responses in later layers) corresponding to the two phases are denoted as $f_1(\textup{X}^\textup{A}_i; \mathbf{\Theta}_1)$ and
$f_2(\textup{X}^\textup{B}_i; \mathbf{\Theta}_2)$, where the two sub-networks are parameterized by $\mathbf{\Theta}_1$ and $\mathbf{\Theta}_2$ respectively. The outputs of two branches will be simultaneously fed to the final classification layer.
In order to better integrate the outcomes from the two branches, we propose to use a pairing loss which exploits the consensus of $f_1(\textup{X}_i^\textup{A}; \mathbf{\Theta}_1)$ and
$f_2(\textup{X}_i^\textup{B}; \mathbf{\Theta}_2)$ during training. The loss is formulated as following:
\begin{equation}
\resizebox{0.92\linewidth}{!}{
$\mathcal{L}_{corr}(\textup{X}^\textup{A}_{i}, \textup{X}^\textup{B}_{i}; \mathbf{\Theta}) = -\frac{\sum_{j=1}^{N} \big( f_1(\textup{X}^\textup{A}_{ij}) - \overline{f_1(\textup{X}^\textup{A}_{i})}  \big) \big( f_2(\textup{X}^\textup{B}_{ij}) - \overline{f_2(\textup{X}^\textup{B}_{i})}  \big)}{ \sqrt{\sum_{j=1}^{N} \big( f_1(\textup{X}^\textup{A}_{ij}) - \overline{f_1(\textup{X}^\textup{A}_{i})}  \big)^2 \sum_{j=1}^{N} \big( f_2(\textup{X}^\textup{B}_{ij}) - \overline{f_2(\textup{X}^\textup{B}_{i})}  \big)^2 }}$,
}
\end{equation}
where $N$ denotes the total number of voxels in the $i$-th sample and $\mathbf{\Theta}$ denotes the parameters of the entire network. During the training stage, we impose this additional loss to further encourage the commonality between the two intermediate outputs. The overall loss is the weighted sum of this additional penalty term and the standard voxel-wise cross-entropy loss:
\begin{equation}
\resizebox{0.85\linewidth}{!}{
$\mathcal{L}_{total} =  -\frac{1}{N} \bigg[\sum\limits_{j=1}^{N} 
\sum\limits_{k=0}^K \mathds{1}  (y_{ij}=k) \log p^k_{ij} \bigg] +\lambda\mathcal{L}_{corr}(\textup{X}^\textup{A}_{i}, \textup{X}^\textup{B}_{i}; \mathbf{\Theta})$,
}
\end{equation}
where $p^k_{ij}$ denotes the probability of the $j$-th voxel be classified as label $k$ on the $i$-th sample and $\mathds{1} (\cdot) $ is the indicator function. $K$ is the total number of classes. The overall objective function is optimized via stochastic gradient descent.

\section{Experiments}
\label{Experiments}

\subsection{Experiment setup}
\label{Experiments:DatasetEvalutation}

\paragraph{Data acquisition.} This is an institutional review board approved HIPAA compliant retrospective case control study. $239$ patients with pathologically proven PDAC  were retrospectively identified from the radiology and pathology databases from 2012 to 2017 and the cases with $\leq$ 4cm tumor (PDAC mass) diameter were selected for the experiment. PDAC patients were scanned on a 64-slice multidetector CT scanner (Sensation 64, Siemens Healthineers) or a dual-source multidetector CT scanner (FLASH, Siemens Healthineers). PDAC patients were injected with 100-120 mL of iohexol (Omnipaque, GE Healthcare) at an injection rate of 4-5 mL/sec. Scan protocols were customized for each patient to minimize dose. Arterial phase imaging was performed with bolus triggering, usually \app30 seconds post-injection, and venous phase imaging was performed \app60 seconds.

\paragraph{Evaluation.}Denote $\mathcal{Y}$ and $\mathcal{Z}$ as the set of foreground voxels in the ground-truth and prediction,
\emph{i.e.}, ${\mathcal{Y}}={\left\{i\mid y_i=1\right\}}$ and ${\mathcal{Z}}={\left\{i\mid z_i=1\right\}}$.
The accuracy of segmentation is evaluated by the Dice-S{\o}rensen coefficient (DSC):
${\mathrm{DSC}\!\left(\mathcal{Y},\mathcal{Z}\right)}=
    {\frac{2\times\left|\mathcal{Y}\cap\mathcal{Z}\right|}{\left|\mathcal{Y}\right|+\left|\mathcal{Z}\right|}}$. 
We evaluate DSCs of all three targets, \emph{i.e.}, abnormal pancreas, PDAC mass and pancreatic duct. All experiments are conducted by three-fold cross-validation, \emph{i.e.}, training the models on two folds and testing them on the remaining one. 
Through our experiment, abnormal pancreas stands for the union of normal pancreatic tissues, PDAC mass and pancreatic duct.
The average DSC of all cases as well as the standard deviations are reported.

\subsection{Implementation details}
\label{sec:imp_details}
Our experiments were performed on the whole CT scan
and the implementations are based on PyTorch. 
We adopt a variation of diffeomorphic demons with direction-dependent regularizations~\cite{Vercauteren2009,reaungamornrat2016mind} for accurate and efficient deformable registration between the two phases. 
For data pre-processing, we truncated the raw intensity values within the range [-100, 240] HU and normalized each raw CT case to have zero mean and unit variance. 
The input sizes of all networks are set as $64\!\times\!64\!\times\!64$. The coefficient of the correlation loss $\lambda$ is set as $0.5$. No further post-processing strategies were applied.

We also used data augmentation during training.
Different from single-phase segmentation which commonly uses rotation and scaling~\cite{Kamnitsas_2016_Efficient,zhu20173d}, virtual sets~\cite{zhang2017mixup} are also utilized in this work. Even though arterial and venous phase scanning are customized for each patient, the level of enhancement can be different from patients by variation of blood circulation, which causes inter-subject enhancement variations on each phase.
Therefore we construct virtual examples by interpolating between venous and arterial data, similar to~\cite{zhang2017mixup}. The $i$-th augmented training sample pair can be written as: 
  $\tilde{\textup{X}}^\textup{A}_i = \lambda \textup{X}^\textup{A}_i + (1 - \lambda) \textup{X}^\textup{B}_i, \quad
  \tilde{\textup{X}}^\textup{B}_i = \lambda \textup{X}^\textup{B}_i + (1 - \lambda) \textup{X}^\textup{A}_i,$
where $\lambda \sim \text{Beta}(\alpha, \alpha) \in [0, 1]$. The final outcome of HPN is obtained by taking the union of predicted regions from models trained with the original paired sets and the virtual paired sets. We set the hyper-parameter $\alpha = 0.4$ following~\cite{zhang2017mixup}.

\begin{table}[t]
\centering
\scriptsize
\setlength{\tabcolsep}{2mm}
\renewcommand\arraystretch{1}
\resizebox{0.9\linewidth}{!}{
\begin{tabular}{l|c |c| c}
\hline
\hline
Method   &Abnormal pancreas  & PDAC mass &pancreatic duct  \\
\hline
3D-UNet-single-phase (Arterial)			&    78.35 $\pm$ 11.89 & 52.40 $\pm$ 27.53             &  38.35 $\pm$ 28.98   \\
3D-UNet-single-phase (Venous)		&    79.61 $\pm$ 10.47 &  53.08 $\pm$ 27.06              &  40.25 $\pm$ 27.89 \\
3D-UNet-multi-phase (fusion)			&    80.05 $\pm$ 10.56 & 52.88 $\pm$ 26.97           &  39.06 $\pm$ 27.33   \\
3D-UNet-multi-phase-HyperNet	&    82.45 $\pm$ 9.98 & 54.36 $\pm$ 26.34            & 43.27 $\pm$ 26.33   \\
3D-UNet-multi-phase-HyperNet-aug		&    83.67 $\pm$ 8.92 & 55.72 $\pm$ 26.01             &  43.53 $\pm$ 25.94   \\
3D-UNet-multi-phase-HPN (Ours)	&    \textbf{84.32} $\pm$ \textbf{8.59} & \textbf{57.10} $\pm$ \textbf{24.76}             &  \textbf{44.93} $\pm$ \textbf{24.88}  \\
\hline
3D-ResDSN-single-phase (Arterial)			&    83.85 $\pm$ 9.43  & 56.21 $\pm$ 26.33             &  47.04 $\pm$ 26.42   \\
3D-ResDSN-single-phase (Venous)			&    84.92 $\pm$ 7.70 & 56.86 $\pm$ 26.67            &  49.81 $\pm$ 26.23   \\
3D-ResDSN-multi-phase (fusion)			&    85.52 $\pm$ 7.84 & 57.59 $\pm$ 26.63           &  48.49 $\pm$ 26.37   \\
3D-ResDSN-multi-phase-HyperNet		&    85.79 $\pm$ 8.86 & 60.87 $\pm$ 24.95             &  54.18 $\pm$ 24.74   \\
3D-ResDSN-multi-phase-HyperNet-aug		&    85.87 $\pm$ 7.91 & 61.69 $\pm$ 23.24             &  54.07 $\pm$ 24.06   \\
3D-ResDSN-multi-HPN (Ours)				& \textbf{86.65} $\pm$ \textbf{7.46} & \textbf{63.94} $\pm$ \textbf{22.74}           &  \textbf{56.77} $\pm$ \textbf{23.33}    \\
\hline
\hline
\end{tabular}
}
\caption{DSC (\%) comparison of abnormal pancreas, PDAC mass and pancreatic duct. We report results in the format of mean $\pm$ standard deviation.}
\label{tbl:results}
\end{table}

\subsection{Results and Discussions}
\label{Experiments:Results}
All results are summarized in Table~\ref{tbl:results}. 
We compare the proposed HPN with the following algorithms: 1) single-phase algorithms which are trained exclusively on one phase (denoted as ``single-phase''); 2) multi-phase algorithm where both arterial and venous data are trained using a dual path network bridged with hyper connections (denoted as ``HyperNet''). 
In general, compared with single-phase algorithms, multi-phase algorithms (\emph{i.e.}, HyperNet, HPN) observe significant improvements for all target structures. It is no surprise to observe such a phenomenon as more useful information is distilled for multi-phase algorithms.  

\paragraph{Efficacy of hyper-connections.} To show the effectiveness of hyper-connections, output from different phases (using single-phase algorithms) are fused by taking at each position the average probability (denoted as ``fusion''). However, we observe that simply fusing the outcomes from different phases usually yield either similar or slightly better performances compared with single-phase algorithms. This indicates that simply fusing the estimations during the inference stage cannot effectively integrate multi-phase information. By contrast, hyper-connections enable the training process to be communicative between the two phase branches and thus can efficiently elevate the performance. Note that directly applying~\cite{dolz2018hyperdense} yield unsatisfactory results. Our hyper-connections are not densely connected but are carefully designed based on previous state-of-the-art on PDAC segmentation~\cite{zhu20173d} for better segmentation of PDAC. Meanwhile, we show much better performance of $63.94\%$ compared to $56.46\%$ reported in~\cite{zhu20173d}. 

\begin{figure}[t!]
\centering
\includegraphics[width=0.85\linewidth]{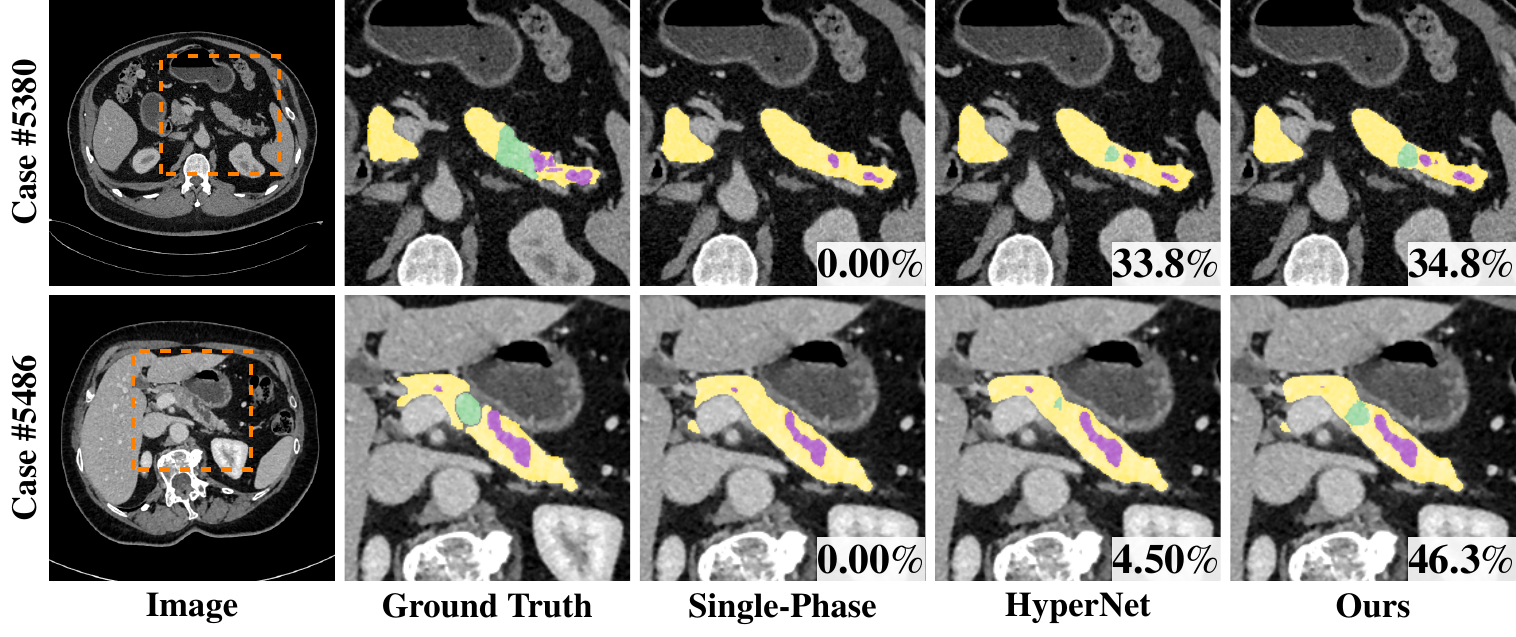}
\caption{Qualitative comparison of different methods, where HPN enhances PDAC mass segmentation (green) significantly compared with other methods. (Best viewed in color)}
\label{fig:qualitative}
\end{figure}

\begin{figure}[t!]
\centering
\includegraphics[width=0.85\linewidth]{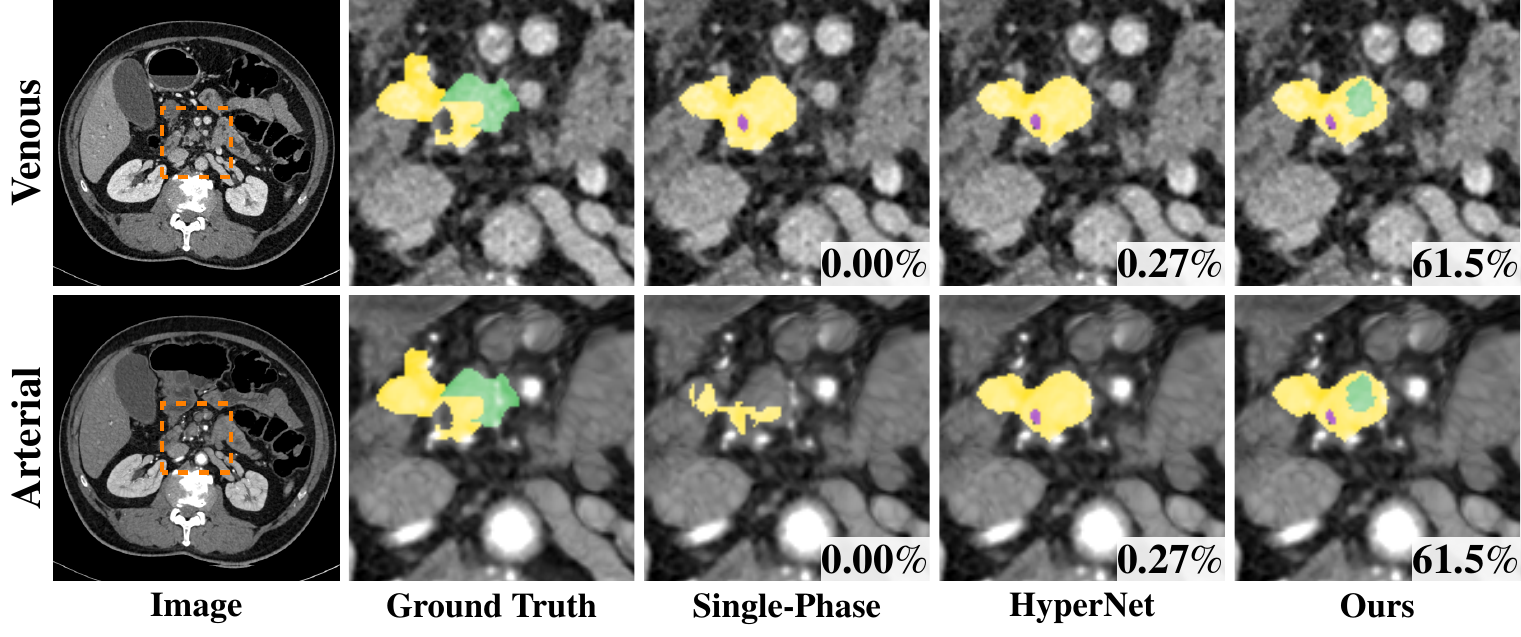}
\caption{Qualitative example where HPN detects the PDAC mass (green) while single-phase methods for both phases fail. From left to right: venous and arterial images (aligned), groundtruth, predictions of single-phase algorithms, HyperNet prediction, HPN prediction (overlayed with venous and arterial images). (Best viewed in color)
}
\label{fig:missing}
\end{figure}

\paragraph{Efficacy of data augmentation.} From Table~\ref{tbl:results}, compared with HyperNet, HyperNet-aug witnesses performance gain especially for PDAC mass (\emph{i.e.}, from $60.87\%$ to $61.69\%$ for 3D-ResDSN; from $54.36\%$ to $55.72\%$ for 3D-UNet), which validates the usefulness of using virtual paired sets as data augmentation.

\paragraph{Efficacy of HPN.} We can observe additional benefit of our HPN over hyperNet-aug (\emph{e.g.}, abnormal pancreas: $85.87\%$  to $86.65\%$, PDAC mass: $61.69\%$ to $63.94\%$, pancreatic duct: $54.07\%$ to $56.77\%$, 3D-ResDSN). Overall, HPN observes an evident improvement compared with HyperNet, \emph{i.e.}, abnormal pancreas: $85.79\%$  to $86.65\%$, PDAC mass: $61.69\%$ to $63.94\%$, pancreatic duct: $54.07\%$ to $56.77\%$ (3D-ResDSN). The \emph{p-value}s for testing significant difference between hyperNet and our HPN of all 3 targets are $p < 0.0001$, which suggests a general statistical improvement. 
We also show two qualitative examples in Fig.~\ref{fig:qualitative}, where HPN shows much better segmentation accuracy especially for PDAC mass.

Another noteworthy fact is that $11/239$ cases are false negatives which failed to detect any PDAC mass using either phase (Dice = $0\%$). Out of these 11 cases, 7 cases are successfully detected by HPN. An example is shown in Fig.~\ref{fig:missing} --- the PDAC mass is missing from both single phases and almost missing in the original HyperNet (DSC=$0.27\%$), but our HPN can detect a reasonable portion of the PDAC mass (DSC=$61.5\%$).

The deformable registration error by computing pancreas surface distances between two phases is $1.01\pm0.52 mm$ (mean $\pm$ standard deviations) which can be considered as acceptable for this study. However, the effects between different alignments can be described as a further study.

\section{Conclusions}
\label{Conclusions}
Motivated by the fact that radiologists usually rely on analyzing multi-phase data for better image interpretations, we develop an end-to-end framework, HPN, for multi-phase image segmentation. 
Specifically, HPN consists of a dual path network where different paths are connected for multi-phase information exchange, and an additional loss is added for removing view divergence.
Extensive experiment results demonstrate that the proposed HPN can substantially and significantly improve the segmentation performance, \emph{i.e.}, HPN reports an improvement up to $7.73\%$ in terms of DSC compared to prior arts which use single phase data. In the future, we plan to examine the behaviour of HPN when using different alignment strategies and try to extend the current approach to other multi-phase learning problems.\\

\noindent {\bf Acknowledgements.}
This work was supported by the Lustgarten Foundation for Pancreatic Cancer Research.
We thank Fengze Liu, Yingda Xia, Qihang Yu and Zhuotun Zhu for instructive discussions.

\bibliographystyle{splncs04}
\bibliography{paper}

\begin{thebibliography}{10}
\providecommand{\url}[1]{\texttt{#1}}
\providecommand{\urlprefix}{URL }
\providecommand{\doi}[1]{https://doi.org/#1}

\bibitem{attiyeh2018survival}
Attiyeh, M.A., Chakraborty, J., Doussot, A., Langdon-Embry, L., Mainarich, S.,
  et~al.: Survival prediction in pancreatic ductal adenocarcinoma by
  quantitative computed tomography image analysis. Annals of surgical oncology
  \textbf{25} (2018)

\bibitem{cciccek20163d}
{\c{C}}i{\c{c}}ek, {\"O}., Abdulkadir, A., Lienkamp, S.S., Brox, T.,
  Ronneberger, O.: 3d u-net: learning dense volumetric segmentation from sparse
  annotation. In: MICCAI. pp. 424--432 (2016)

\bibitem{dolz2018hyperdense}
Dolz, J., Gopinath, K., Yuan, J., Lombaert, H., Desrosiers, C., Ayed, I.B.:
  Hyperdense-net: A hyper-densely connected cnn for multi-modal image
  segmentation. TMI  (2018)

\bibitem{dou2016automatic}
Dou, Q., Chen, H., Yu, L., Zhao, L., Qin, J., Wang, D., Mok, V.C., Shi, L.,
  Heng, P.A.: Automatic detection of cerebral microbleeds from mr images via 3d
  convolutional neural networks. TMI  \textbf{35}(5),  1182--1195 (2016)

\bibitem{Kamnitsas_2016_Efficient}
Kamnitsas, K., Ledig, C., Newcombe, V., Simpson, J., Kane, A., Menon, D.,
  Rueckert, D., Glocker, B.: {Efficient Multi-Scale 3D CNN with Fully Connected
  CRF for Accurate Brain Lesion Segmentation}. arXiv

\bibitem{li2019multimodal}
Li, Y., Liu, J., Gao, X., Jie, B., Kim, M., Yap, P.T., Wee, C.Y., Shen, D.:
  Multimodal hyper-connectivity of functional networks using
  functionally-weighted lasso for mci classification. Medical image analysis
  \textbf{52},  80--96 (2019)

\bibitem{Long_2015_Fully}
Long, J., Shelhamer, E., Darrell, T.: {Fully Convolutional Networks for
  Semantic Segmentation}. In: CVPR (2015)

\bibitem{Milletari_2016_VNet}
Milletari, F., Navab, N., Ahmadi, S.: {V-Net: Fully Convolutional Neural
  Networks for Volumetric Medical Image Segmentation}. In: 3DV (2016)

\bibitem{reaungamornrat2016mind}
Reaungamornrat, S., De~Silva, T., Uneri, A., Vogt, S., Kleinszig, G., Khanna,
  A.J., Wolinsky, J.P., Prince, J.L., Siewerdsen, J.H.: Mind demons: symmetric
  diffeomorphic deformable registration of mr and ct for image-guided spine
  surgery. TMI  \textbf{35}(11),  2413--2424 (2016)

\bibitem{Ronneberger_2015_UNet}
Ronneberger, O., Fischer, P., Brox, T.: {U-Net: Convolutional Networks for
  Biomedical Image Segmentation}. In: MICCAI (2015)

\bibitem{Roth_2016_Spatial}
Roth, H., Lu, L., Farag, A., Sohn, A., Summers, R.: {Spatial Aggregation of
  Holistically-Nested Networks for Automated Pancreas Segmentation}. In: MICCAI
  (2016)

\bibitem{Vercauteren2009}
Vercauteren, T., Pennec, X., Perchange, A., Ayache, N.: Diffeomorphic demons:
  efficient non-parametric image registration. NeuroImage  \textbf{45}(1),
  S61--S82 (2009)

\bibitem{wang2019abdominal}
Wang, Y., Zhou, Y., Shen, W., Park, S., Fishman, E.K., Yuille, A.L.: Abdominal
  multi-organ segmentation with organ-attention networks and statistical
  fusion. Medical image analysis  \textbf{55},  88--102 (2019)

\bibitem{yao2017deep}
Yao, J., Zhu, X., Zhu, F., Huang, J.: Deep correlational learning for survival
  prediction from multi-modality data. In: MICCAI. pp. 406--414 (2017)

\bibitem{zhang2017mixup}
Zhang, H., Cisse, M., Dauphin, Y.N., Lopez-Paz, D.: mixup: Beyond empirical
  risk minimization. In: ICLR (2018)

\bibitem{zhang2015deep}
Zhang, W., Li, R., Deng, H., Wang, L., Lin, W., Ji, S., Shen, D.: Deep
  convolutional neural networks for multi-modality isointense infant brain
  image segmentation. NeuroImage  \textbf{108},  214--224 (2015)

\bibitem{zhou2017deep}
Zhou, Y., Xie, L., Fishman, E.K., Yuille, A.L.: Deep supervision for pancreatic
  cyst segmentation in abdominal ct scans. In: MICCAI. pp. 222--230 (2017)

\bibitem{zhou2017fixed}
Zhou, Y., Xie, L., Shen, W., Wang, Y., Fishman, E.K., Yuille, A.L.: A
  fixed-point model for pancreas segmentation in abdominal ct scans. In:
  MICCAI. pp. 693--701 (2017)

\bibitem{zhu2019anatomynet}
Zhu, W., Huang, Y., Zeng, L., Chen, X., Liu, Y., Qian, Z., Du, N., Fan, W.,
  Xie, X.: Anatomynet: Deep learning for fast and fully automated whole-volume
  segmentation of head and neck anatomy. Medical physics  \textbf{46}(2),
  576--589 (2019)

\bibitem{zhu20173d}
Zhu, Z., Xia, Y., Xie, L., Fishman, E.K., Yuille, A.L.: Multi-scale
  coarse-to-fine segmentation for screening pancreatic ductal adenocarcinoma.
  arXiv  (2018)

\end{thebibliography}
\end{document}